# Valence band and core level photoemission spectroscopy of LaFeAsO$_{1-x}$F$_x$

A. Koitzsch[1], D. Inosov[1], J. Fink[1,2], M. Knupfer[1], H. Eschrig[1], S. V. Borisenko[1], G. Behr[1], A. Köhler[1], J. Werner[1], B. Büchner[1], R. Follath[2], and H. A. Dürr[2]

[1] *Institute for Solid State Research, IFW Dresden, P.O. Box 270116, D-01171 Dresden, Germany.*

[2] *BESSY GmbH, Albert-Einstein-Strasse 15, 12489 Berlin, Germany*

**Abstract**: We have investigated the electronic structure of LaFeAsO$_{1-x}$F$_x$ (x = 0; 0.1; 0.2) by angle-integrated photoemission spectroscopy and local density approximation (LDA) based band structure calculations. The valence band consists of a low energy peak at E ≈ -0.25 eV and a broad structure around E ≈ -5 eV in qualitative agreement with LDA. From the photon energy dependence of these peaks we conclude that the former derives almost exclusively from Fe 3d states. This constitutes experimental evidence for the strong iron character of the relevant states in a broad window around $E_F$ and confirms theoretical predictions.

The recent discovery of the superconducting oxypnictides has sparked immediate and intense scientific effort. Naturally the question about the origin of superconductivity and why exactly it reaches such high $T_c$ is the driving force of the field, especially since the same problem for the cuprate high temperature superconductors still awaits solution. Here superconductivity is found in yet another tetragonal highly two-dimensional system. Therefore the new materials offer a fresh perspective on the general problem of high $T_c$ and, possibly, insight into so far elusive aspects.

Superconducting transition temperatures up to $T_c = 43$ K have been reached for $LaFeAsO_{1-x}F_x$ [1, 2]. Substitution of La by other rare earth elements enhances $T_c$ even further to values beyond 50 K [3, 4]. As in the cuprate superconductors and in some heavy fermion compounds superconductivity is found in proximity to magnetic ordering. For undoped LaFeAsO a commensurate antiferromagnetic phase is found below 140 K [5, 6]. Several studies present experimental evidence for line nodes of the superconducting order parameter [7, 8, 9] but also s-wave pairing seems possible [10, 11]. The electronic structure and its implications for superconductivity have been investigated from the theoretical side by [12, 13, 14, 15, 16, 17]. $LaFeAsO_{1-x}F_x$ crystals consist of alternating $LaO_{1-x}F_x$ and FeAs layers. It is assumed that the $LaO_{1-x}F_x$ layers serve as ionic charge reservoirs for the covalently bound metallic FeAs layers where the superconductivity appears. According to band structure calculations the layered crystal structure translates into a highly two-dimensional electronic structure with cylindrical Fermi surface sheets. The density of states in the vicinity of the Fermi energy is dominated by iron character. Previous photoemission studies found gross agreement between experiment and theory [18, 19, 10].

Here we report on angle integrated photoemission measurements of $LaFeAsO_{1-x}F_x$ ($x = 0, 0.1, 0.2$). We have measured the photointensity of polycrystalline material for various excitation energies ranging from $h\nu = 15$ eV to $h\nu = 200$ eV at $T = 30$ K. Experimentally a peak near the Fermi energy and a second broad structure several eV below $E_F$ is observed for all doping levels. From the excitation energy dependence we conclude that the near $E_F$ peak consists predominantly of iron states. We find doping dependent quantities in the valence and core level spectra. The energy position of the lines shift generally to higher energies with electron doping.

Polycrystalline samples of $LaFeAsO_{1-x}F_x$ were prepared by using a two-step solid state reaction method, similar to that described by Zhu et al. [20]. The samples consist of 1 to 100 μm sized grains of $LaFeAsO_{1-x}F_x$. The crystal structure and the composition were investigated by powder X-ray diffraction (XRD) and wavelength dispersive X-ray spectroscopy (WDX).

Critical temperatures of $T_c \approx 23$ K and $T_c \approx 10$ K for x = 0.1 and x = 0.2, respectively, have been extracted from magnetization and resistivity measurements. The undoped sample shows a transition to a commensurate spin density wave below $T_N$ = 138 K [6]. The XRD analysis shows a phase purity of 96%, 92% and 89% for x = 0, x = 0.1 and x = 0.2 respectively.
The data were measured using synchrotron radiation at the "1$^3$" ARPES station with a Scienta R4000 spectrometer at BESSY. The energy resolution was better than 25 meV for excitation energies below 100 eV and better than 40 meV for excitation energies up to 200 eV. The measurements have been performed at temperatures below 30 K. The samples have been scraped *in situ* before measurements at a pressure of p = 1 · 10$^{-7}$ mbar. The base pressure in the measurement chamber was p = 1 · 10$^{-10}$ mbar.

Figure 1 presents the valence band of undoped LaFeAsO taken with different photon energies. For reasons explained below the spectra have been normalized to the high energy shoulder of the broad peak centered at about E ≈ -5 eV (marked by the arrow). We compare the experimental data to LDA based orbital resolved density of states (DOS) calculations in panel (b). The main features of the valence band for all photon energies are a peak near the Fermi energy at E ≈ -0.25 eV and the broad peak around E ≈ -5 eV [21]. Note that the peak near $E_F$ is not cut off by the Fermi edge step. The inset of panel (a) shows a zoom of the vicinity of $E_F$ for hν = 15 eV. The Fermi edge appears as a small slope change at the low energy tail of the spectral weight near $E_F$, which is clearly visible at low temperatures only. In between the near $E_F$ peak and the broad peak a plateau-like region is observed with a small peak at E ≈ 1.7 eV. The broad peak has a complex structure and consists at least of two separate features. The low energy peak and the broad peak have a significant dependence on the photon energy. The intensity of the low energy peak increases drastically with increasing photon energy relative to the broad peak. The center of gravity of the broad peak shifts towards lower energies due to the intensity increase of the low energy shoulder, which becomes more intense than the high energy shoulder at hν = 95 eV. The presented spectra are taken at T = 30 K, but no major changes are observed when crossing the magnetic ordering temperature $T_N$ = 138 K.
The reason for the hν dependent intensity variations lies in the hν dependence of the photoemission cross section. The opposite behavior of the high energy shoulder of the broad peak and the low energy peak suggests that they arise from different atomic orbitals. We show the energy dependence of the cross section of the potentially important valence orbitals, namely Fe 3d, As 4p, and O 2p as a function of photon energy in Fig 1c. As 4p is important for low energies (< 25 eV) only. For energies above ~ 25 eV the spectra will be governed by Fe 3d and O 2p emission. For increasing photon energy Fe 3d dominates. Fig 1d shows the ratio of the cross sections Fe 3d/O 2p (blue line). It increases monotonically in the measured range and reaches a value of 6.5 for hν = 200 eV. Motivated by the increase of the low energy peak and the decrease of the high energy shoulder with increasing hν we assume, that the former is due to Fe 3d and the latter is mainly due to O 2p. We evaluate the intensity of the low energy peak by taking the integral from zero to E = -1 eV without any background treatment (red box). Since we normalized the spectra to the high energy shoulder this integral value corresponds automatically to the experimental intensity ratio. Those values are plotted in Fig. 1d as red points. We find satisfactory agreement and conclude *a posteriori* the correctness of our assumptions.

The LDA calculations are in qualitative agreement with these findings. The O 2p orbital DOS is found at the high energy side of the broad structure, the low energy peak is due to Fe 3d, and is offset from the Fermi energy. Also there is substantial Fe 3d weight in the low energy side of the broad peak in agreement with the increasing intensity with increasing photon energy. The small peak at E = -1.7 eV and the plateau can be associated with a small peak in the DOS at the same energy. However, there are also significant discrepancies: compared to theory the energy position of the broad peak is shifted to higher energies. Also, the width of the low energy peak is smaller in experiment than in theory. According to theory at low photon energies the As 4p states should play a more prominent role which is not obvious in the data. Although in Fig 1 only data for undoped LaFeAsO are shown we would like to point out that at all doping levels a similar behavior is observed.

In Fig. 2 we investigate the doping dependence of the valence band. Electron doping is achieved by substituting oxygen by fluor atoms in the LaO layers. The LaO layers are considered as chemically inert charge reservoirs. The doping level x corresponds in this picture to the number of electrons injected to the FeAs layers per formula unit. We show the doping dependence of the valence band for four different excitation energies in Fig. 2(a)-(d). The spectra are in each case normalized to the number of available electrons in the given energy window weighted by their cross section. [23] We have confirmed that normalization to the La 4d or the La 5p lines gives qualitatively equivalent results. All doping levels show the same structure of the valence band for all excitation energies: the low energy peak is followed by a plateau of nearly constant intensity and eventually an intensity increase due to the broad peak feature. For hν = 200 eV (panel (a)) an increase of the plateau intensity is observed with electron doping. For hν = 150 eV excitation energy the results (panel (b)) are similar to hν = 200 eV. Within the studied doping range, the valence band does not dramatically change but pre-serves its overall structure. The inset of panel (c) shows a comparison of the low energy peak for x = 0.2 F- doping and x = 0. The other doping level has been skipped for clarity. We find a small shift of the peak for electron doping towards higher energies of $\Delta E \approx 30$ meV. The same trend is found for hν = 48 eV (inset panel d) and for the other photon energies. When lowering the excitation energy the spectra become more sensitive to oxygen emission. For hν = 48 eV (panel (d)) the low energy peak is small but the overall doping dependence is consistent. Additionally we find a shift of the broad peak towards larger energies for electron doping.

To investigate possible lineshifts in more detail we present the spin-orbit split La 4d and As 3d core levels in Fig. 3 measured for different doping levels. The vertical bars indicate the measured peak positions. The peakshifts shown in the inset are referenced to zero doping and refer to the $d_{5/2}$ components at the lower energy side. The data show a consistent shift towards higher energies for electron doping. The magnitude of the shift is smaller for the As line. The shape of the peaks changes slightly with doping. This seems particularly true for the high binding energy side and is possibly reflected in the $d_{3/2}$ intensity variations. With doping the chemical environment of La sites changes in the lattice, therefore some changes are expected. The peakshifts indicate a changing chemical potential with doping. Within a rigid band model electron doping would fill so far unoccupied states, thereby shifting the chemical potential "to the right". This effectively increases the binding energy of all electronic states then. This is the observed effect, the energy increases with electron doping. But the change of the binding energy of a specific core level depends not only on the chemical potential shift but also on the specific chemical surroundings of the considered orbital; a possible change of the valence of the atom, the charge balance expressed by the Madelung potential and possibly screening terms [24]. Having this in mind it is not surprising that the La level shifts more than the As level. The ionic $LaO_{1-x}F_x$ layers are not influenced by valence changes or screening. Additional electrons in the FeAs layers, however, tend to counteract shifts towards higher binding energies. Therefore the observed core level shift of the La 4d line of $\Delta E = 200$ meV for x = 0.2 reveals the change of the chemical potential with doping. This interpretation is consistent with our observations for the valence band: the low energy feature, which consists of iron states, shows only a small shift. The oxygen emission, at the other hand, shows again a clear shift with doping, as visible for the hν = 48 eV spectra. For higher photon energies the Fe states dominate also for the broad peak and the shift is not clear anymore.

Our experimental observations show that the oxypnictides behave rather like conventional metals than strongly correlated ones: the density of states and the orbital character agrees qualitatively with LDA calculations, and the energy shift of the levels follows band filling ar-guments. This suggests that correlation effects do not play a dominant role for these materials which clearly distinguishes them from the cuprates. Nevertheless they might not be fully negligible due to the obeserved band narrowing (Fig. 1). Although both materials, cuprates and oxypnictides, have a layered structure with the transition metal ion on a tetragonal lattice we find pure iron states at $E_F$ in the oxypnictides instead of strongly hybridized transition metal ion – ligand states as in the cuprates.

In summary we have investigated the electronic structure of $LaFeAsO_{1-x}F_x$ by angle integrated photoemission spectroscopy and LDA based bandstructure calculations. From the comparison of the valence band with the theoretical density of states we find qualitative agreement. From the photon energy dependence of the spectral weight we explicitly confirm that the low energy density of states consists predominantly of Fe 3d states. We measured the As 3d and La 4d core levels and find a shift of the chemical potential with electron doping of $\Delta E \approx 200$ meV for x = 0.2. The entirety of our results suggests that the oxypnictides are not in a strongly correlated regime.

We acknowledge technical assistance by R. Schönfelder and R. Hübel and financial support by the DFG (SFB 463).

**Captions**

**Fig.1**: (Color online) (a) hv dependent photoemission valence band spectra of LaOFeAs. The arrow marks the point of normalization. The red rectangle is the integration window for the low energy weight shown in panel (d); the inset shows the near $E_F$ region for hv = 15 eV. Note the small change at E = 0 which indicates the Fermi edge. (b) LDA derived orbital resolved. (c) Atomic photoemission cross section for the relevant orbitals [22]. (d) Ratio of the Fe 3d and O 2p cross section from (c) (line) compared to experimental values obtained by integrating the low energy peak (squares).

**Fig. 2**: (Color online) Doping dependence of the valence band for hv = 200 eV (a), 150 eV (b), 95 eV (c), 48 eV (d). The spectra have been normalized to their electron count weighted by the photoemission cross section [Note 1]. Insets show expanded views on the low energy region.

**Fig. 3**: (Color online) Doping dependent spectra of $LaO_{1-x}F_xFeAs$ with x = 0; 0.1; 0.2 for (a) La 4d (b) As 3d. The core levels have been normalized to their integrated intensity. The bars mark the position of the peak maxima. The inset shows the observed peak shifts for La $4d_{5/2}$ (red squares) and As $3d_{5/2}$ (green circles). The spectra are offset vertically for better visibility.

**Fig. 1**

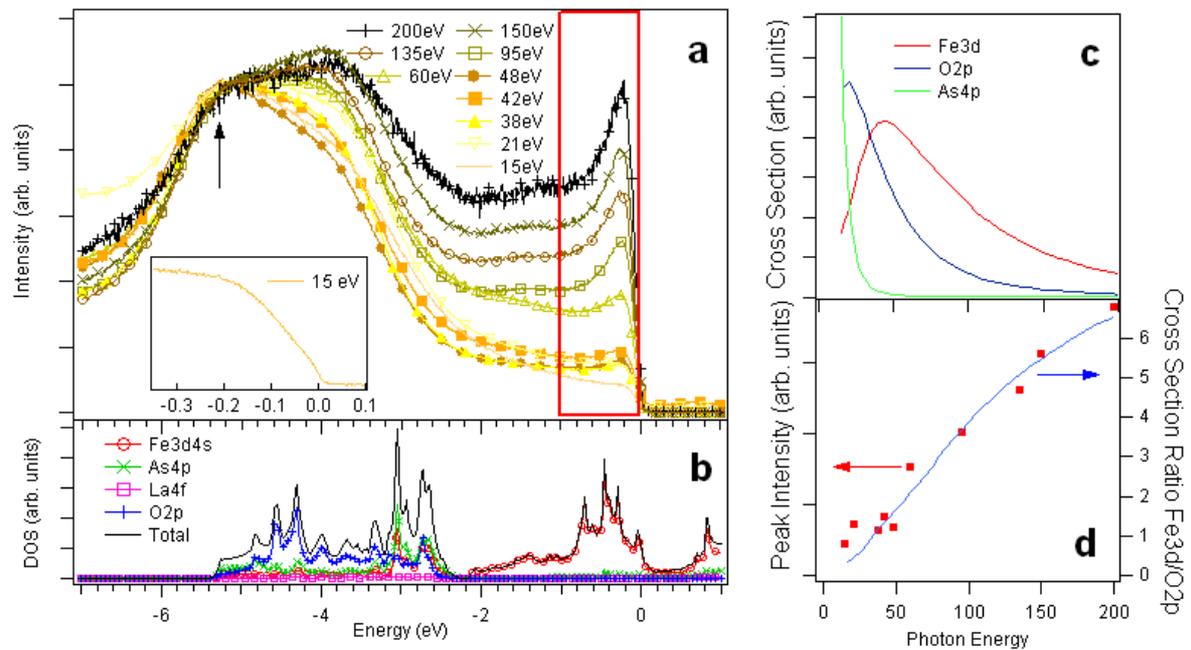

**Fig. 2**

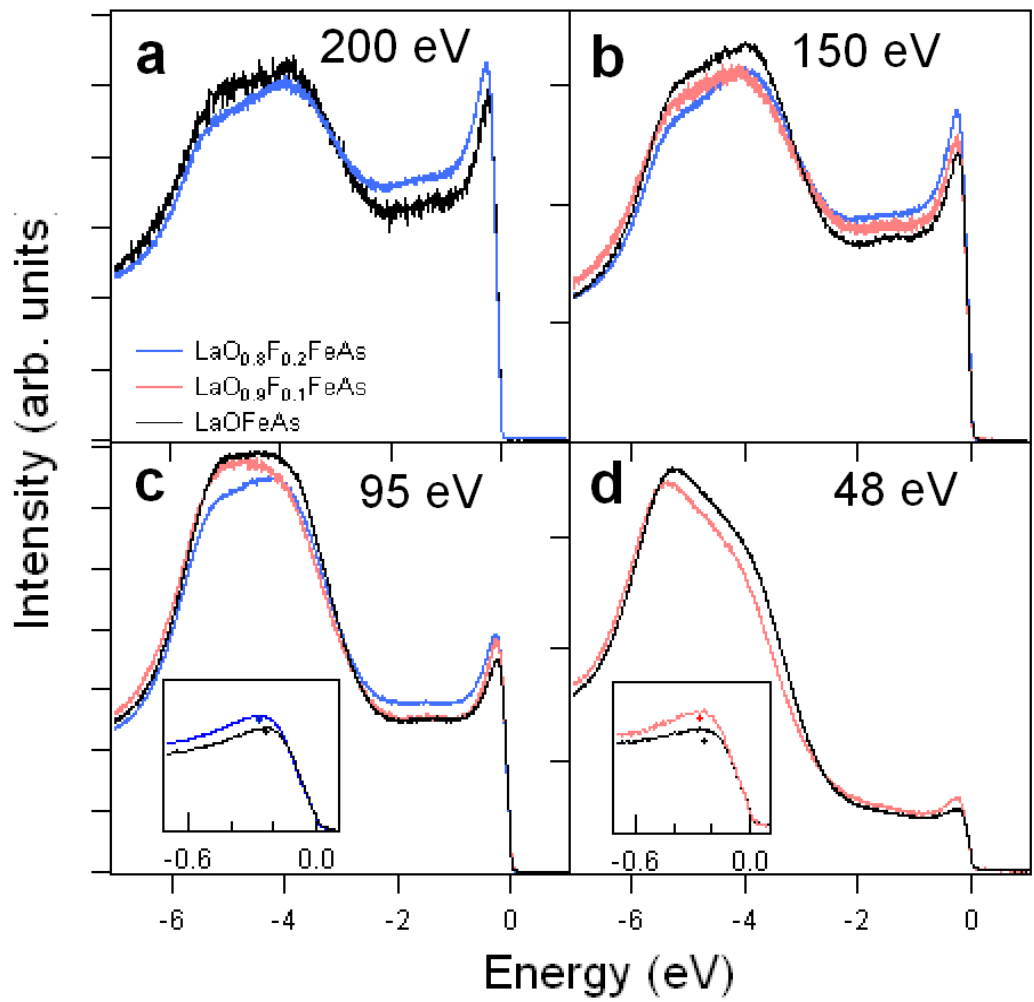

**Fig. 3**

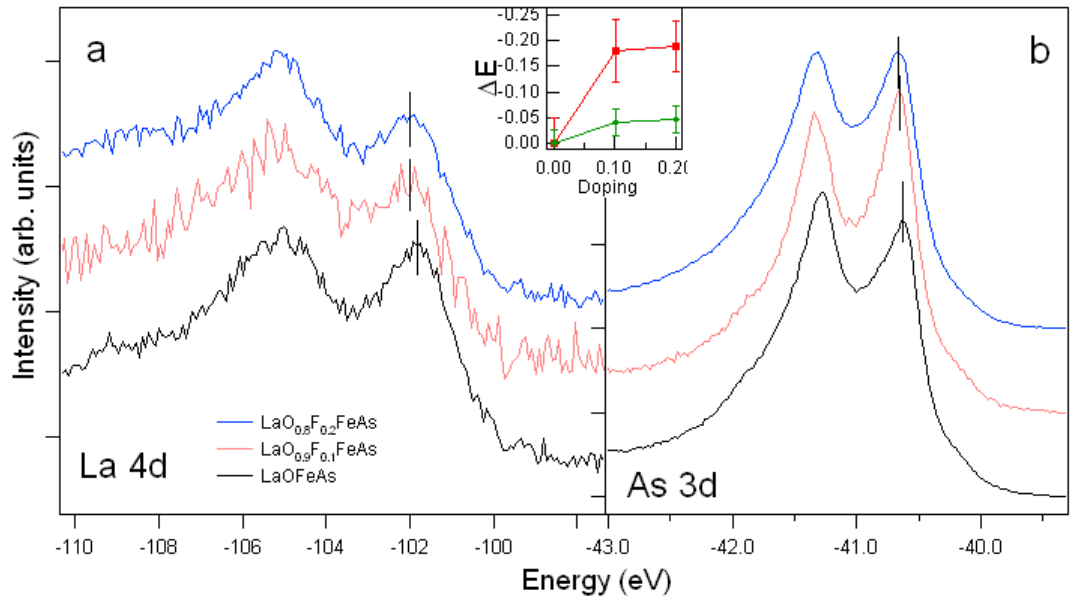